\documentstyle[12pt,epsfig,epsf,twoside,fleqn,espcrc]{article}

\newcommand{\AmS}{{\protect\the\textfont2
  A\kern-.1667em\lower.5ex\hbox{M}\kern-.125emS}}

\title{Excitation Functions of Stopping Power and Flow in Relativistic 
Heavy-Ion Collisions} 

\author{Bao-An Li and C. M. Ko}

\address{Cyclotron Institute and Physics 
Department, Texas A\&M University, College Station, TX 77843, USA}

\begin{document}

\maketitle

\begin{abstract}
Using a relativistic transport (ART) model, we study the stopping power, 
the formation of superdense hadronic matter as well as the strength of 
transverse and radial flow in central Au+Au collisions at beam momentum
from 2 to 12 GeV/c per nucleon. We find that complete stopping is 
achieved in the whole beam momentum range. In particular, the proton 
rapidity distribution scaled by the beam rapidity is independent of 
the beam momentum, and this is in agreement with the experimental findings.  
Also, a large volume of superdense hadronic matter with a local energy 
density exceeding that expected for the transition of a hadronic matter 
to the quark-gluon plasma is formed in collisions at beam momenta greater
than 8 GeV/c per nucleon. Furthermore, it is found that the transverse flow
in these collisions is sensitive to the nuclear equation of state and 
decreases with increasing beam momentum. On the other hand, the radial 
flow is insensitive to the equation of state, and its strength increases
with beam momentum.
\end{abstract}

\section{INTRODUCTION}

Based on hydrodynamical models it has been predicted that the hadronic matter 
to quark-gluon plasma (QGP) phase transition may occur in heavy-ion collisions 
at incident energies between 2 and 10 GeV per nucleon \cite{gyu}. The formation 
of the quark-gluon plasma is expected to be accompanied also by the restoration
of the chiral symmetry. To create the QGP requires that a large volume 
of high energy density matter is formed in the reactions. Whether this
can be achieved in heavy ion collisions depends on the stopping power between
the two colliding nuclei. It is therefore of interest to study how the
stopping power and the volume of high density hadronic matter change with
the incident energy.  Also, collective nuclear flows are possible
signatures for the formation of QGP.  In Ref. \cite{rischke}, it has been 
shown using the hydrodynamical model that formation of the QGP in heavy 
ion collisions can lead to an abrupt change of the transverse and radial flows
with respect to the incident energy. In this talk, we shall report results 
from our study on these questions using A Relativistic Transport (ART) model 
\cite{art1}.  We note that experiments (e.g., E895, E866, E917, etc) are 
being carried out at Brookhaven's AGS to study Au+Au collisions at incident 
beam momenta from 2 to 12 GeV/c per nucleon. Our study is thus useful in 
the search for the signatures of QGP formation and/or chiral symmetry 
restoration in these collisions.

\section{THE EXCITATION FUNCTION OF NUCLEAR STOPPING POWER}

Nuclear stopping power can be inferred from the rapidity distribution
of final nucleons. Complete stopping would lead to a rapidity distribution
which peaks at the mid-rapidity, which has been observed in earlier
experiments at AGS for central collisions of Au+Au at an incident beam
momentum of 12 GeV/c per nucleon \cite{e866n}. Such a proton distribution 
can be well reproduced by our ART model \cite{art1}. This model further 
predicts that complete stopping can also be achieved at lower beam 
momenta \cite{liko96}.  It is, however, surprising to see that 
the scaled proton rapidity distribution, i.e., expressed in terms
of $y/y_{\rm beam}$ where $y_{\rm beam}$ is the beam rapidity, 
from the model calculation at different beam energies all lie on a 
universal curve as shown in Fig. 1. This simple scaling 
is in quantitative agreement with what has been found by Harris \cite{harris} 
based on the analysis of experimental data from heavy ion collisions
at incident energies from 0.25 to 160 GeV per nucleon. This scaling 
behavior may be related to the Fermi-Landau scaling \cite{fl} of 
the pion multiplicity in a nucleon-nucleon collision, i.e., it increases 
more or less linearly with the available center-of-mass energy.
Since the nucleon-nucleon inelastic cross section used in the ART 
model has the Fermi-Landau scaling,
the pion multiplicity per participant in heavy ion collisions 
is also found to follow a similar dependence on the center-of-mass energy 
\cite{liko96}. It is thus of interest to study in the ART model how the 
pion multiplicity and the scaled proton rapidity distribution are modified 
when the Fermi-Landau scaling is violated in the input cross sections.

\vspace*{10cm}
\includegraphics{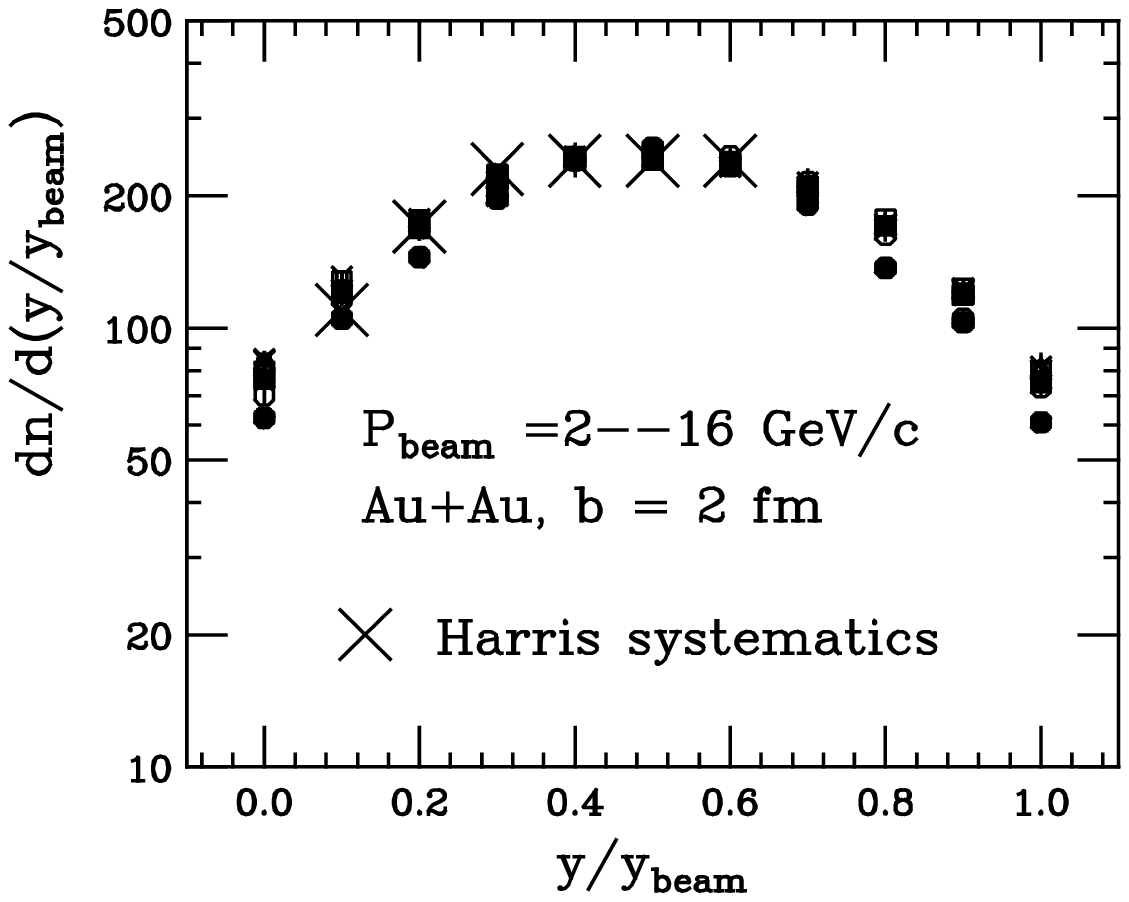}\label{bali-fig1}
\begin{minipage}[t]{6in}
\noindent Figure 1. 
The scaled proton rapidity distribution from the reaction of Au+Au at
an impact parameter of 2 fm and beam momentum of 2 to 16 GeV/c per nucleon. 
The large crosses are from Ref. \protect\cite{harris}.
\end{minipage}
\vskip 1cm

\section{THE EXCITATION FUNCTION OF SUPERDENSE HADRONIC MATTER FORMATION}

The complete stopping of two colliding nuclei in central Au+Au reactions 
at $P_{\rm beam}/A=$ 2 to 12 GeV implies that high energy density 
matter is created in the collision. As a useful guidance for the search 
of QGP at the AGS, we show in Fig. 2 the time evolution of 
the volume of hadronic matter where the local energy density is higher than 
the estimated transition energy density of about 2.5 GeV/fm$^3$ 
between the hadronic matter and the quark-gluon plasma \cite{wong95}.
Our results indicate that the transition to QGP can already occur at 
a beam momentum of about 8 GeV/c. With increasing beam momentum, the
volume of such a high energy density matter becomes more significantly 
larger. 

\vspace*{8.8cm}
\includegraphics{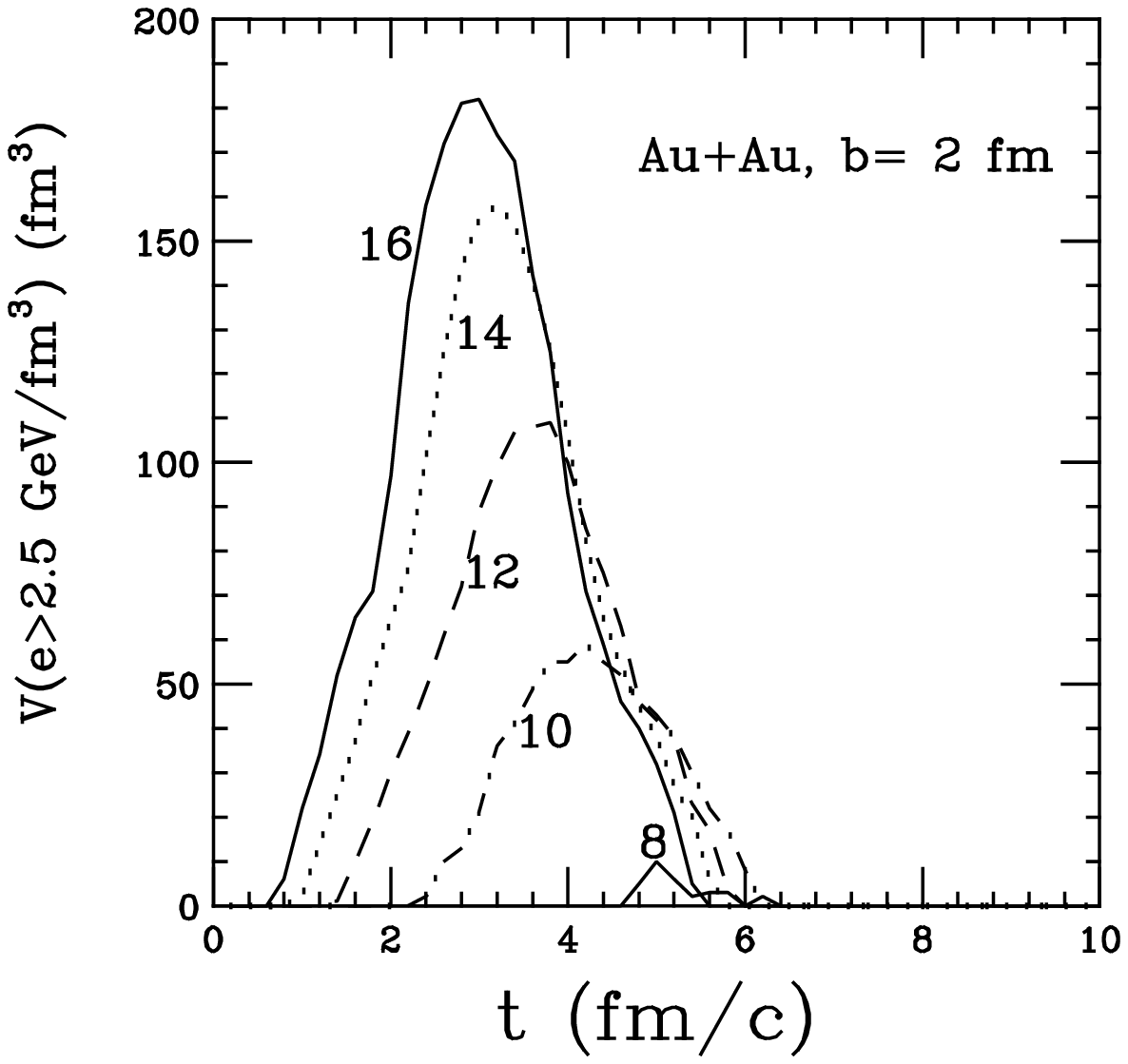}\label{bali-fig2}
\begin{minipage}[t]{6in}
\noindent Figure. 2. 
The volume of hadronic matter with local energy density higher than 
2.5 GeV/fm$^3$ as a function of time and beam momentum for the reaction 
of Au+Au at an impact parameter of 2 fm using the cascade mode of ART.
\end{minipage}
\vskip 1cm

The maximum compression reached in a heavy-ion collision 
can also be determined by solving the Rankine-Hugoniot (RH)
equation \cite{taub}.  In the relativistic hydrodynamical model 
\cite{dan}, the maximum baryon density $\rho$ is found to satisfy 
the following equation,
\begin{equation}\label{shock}
f(\rho)(n-\gamma)-e_0n(B\gamma^2+\gamma^2-1-B\gamma n)=0,
\end{equation}    
where $n \equiv \rho/\rho_0$ and $e_0$ is the energy density of the nuclear
matter in the front of the shock wave; $\gamma$ is the Lorentz factor of the 
beam energy in the nucleus-nucleus center of mass frame; $B$ is the 
ratio of the thermal pressure to the energy density. We find that 
up to the beam momentum of about 10 GeV/c per nucleon a 
value of $B=2/3$ corresponding to a non-relativistic gas can best 
describe our results from the transport model. In the above, 
$f(\rho)$ is defined as
\begin{equation}
f(\rho)\equiv \rho^2d(\frac{e_0(\rho)}{\rho})/d\rho-Be_0(\rho),
\end{equation}
where $e_0(\rho)$ is the energy density at zero temperature. 
For a soft equation of state as used in the transport model, we have
\begin{equation}
e_0(\rho)/\rho=m_N+\frac{a}{2}\frac{\rho}{\rho_0}+\frac{b}{1+\sigma}(\frac{\rho}
{\rho_0})^{\sigma}+\frac{3}{5}E_f(\frac{\rho}{\rho_0})^{2/3}
\end{equation}
and 
\begin{equation}
f(\rho)=\rho_0\left(-\frac{2}{3}m_Nn+\frac{a}{6}n^2+\frac{b(\sigma-2/3)}
{1+\sigma}n^{\sigma+1}\right),
\end{equation}
where $a=-358.1$ MeV, $b=304.8$ MeV, $\sigma=7/6$,
$E_f$ is the Fermi energy in the normal nuclear matter, and $m_N$ is the 
nucleon mass. In the case of a free nucleon gas, i.e., $a=b=0$, one obtains
the following analytical solution for the maximum baryon density \cite{nix},
\begin{equation}\label{max}
n=\frac{5\gamma^2-2\gamma-3}{2(\gamma-1)}
=\frac{1}{2}(5\gamma+3).
\end{equation}

\vspace*{9.5cm}
\includegraphics{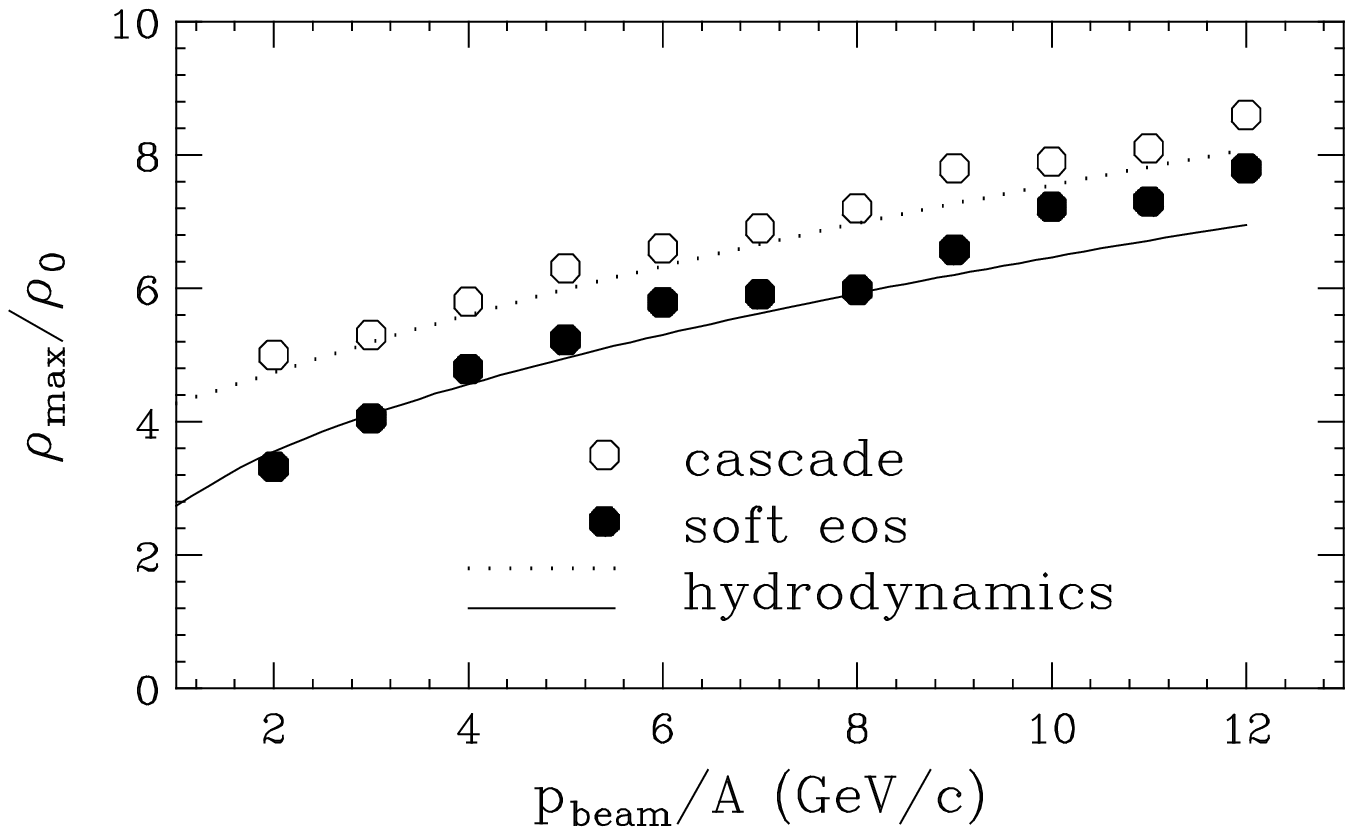}\label{bali-fig3}
\begin{minipage}[t]{6in}
\noindent Figure 3. 
The beam momentum dependence of the maximum 
baryon density for the reaction of Au+Au at an 
impact parameter of 2 fm. The dotted and solid lines 
are predictions of the relativistic hydrodynamics.
\end{minipage}
\vskip 1cm

The maximum baryon density obtained from Eq. (\ref{max}) is shown by 
the dotted line in Fig. 3. The solid line is the result 
from solving numerically Eq. (\ref{shock}) using an iterative method 
and a soft equation of state corresponding to a compressibility of 210 MeV. 
Results from the cascade model and the transport model with the soft 
equation of state are shown by the open and the filled circles, respectively. 
It is amazing that predictions from the two models 
agree very well in both cases of free and interacting nucleon gas,
especially at beam momenta below 9 GeV/c per nucleon. 
Small discrepancies between the two model predictions appear when 
the beam momentum is higher than about 10 GeV/c per nucleon.
This is, however, expected as the system under 
study deviates from a non-relativistic gas at increasingly higher energies, 
and consequently the parameter $B$ should be smaller than 2/3. 
Also, the conditions for the applicability of the hydrodynamical approach,
such as local thermal equilibrium, may not be realized
in the transport model when the incident energy is high.

\vspace*{9.5cm}
\includegraphics{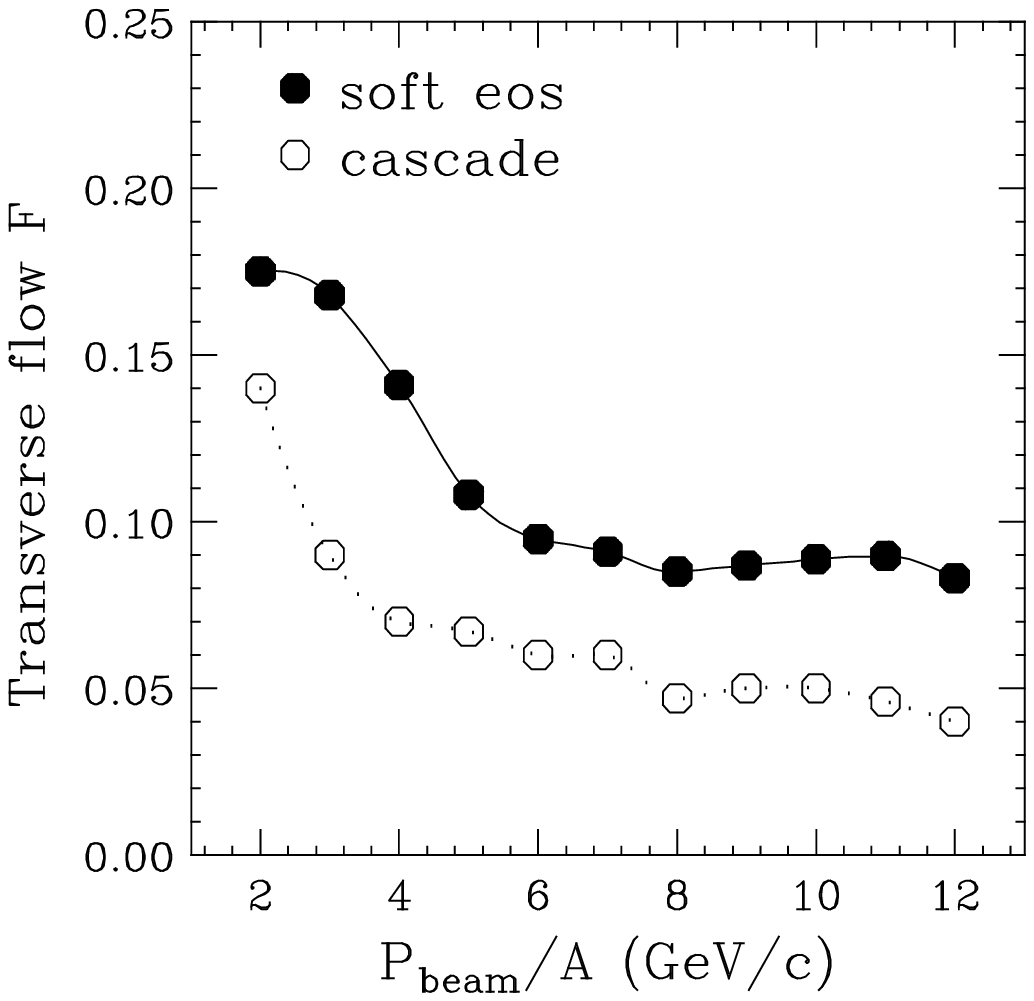}\label{bali-fig4}
\begin{minipage}[t]{6in}
\noindent Figure 4. 
The transverse flow parameter as a function
of beam momentum for the reaction of Au+Au at an impact
parameter of 2 fm.
\end{minipage}
\vskip 1cm

\section{EXCITATION FUNCTIONS OF TRANSVERSE AND RADIAL FLOWS}

To study the transverse flow we use the standard 
method of Danielewicz and Odyniec \cite{dani}, i.e., 
we analyze the average transverse momentum in the reaction 
plane as a function of rapidity. We also extract the flow parameter
$F\equiv (dp_x/dy)_{y=0}$ and study its dependence on the beam energy.
Results of this analysis from both the cascade and the transport
model with the soft equation of state are shown in Fig. \ref{bali-fig4} 
for the reaction of Au+Au at an impact parameter of 2 fm. 
Two interesting features are seen in these results. First, in both cases 
the transverse flow parameter $F$ shows a strong decrease as the beam momentum 
increases from 2 to about 7 GeV/c per nucleon. Above these momenta, it
becomes more or less a constant. Secondly, at all energies the predicted 
flow parameter from the cascade is about a factor of two smaller than that 
using the soft equation of state. These effects
can be understood schematically by considering the maximum flow parameter 
$F_{\rm max}$ as a function of beam momentum, which is approximately given
by 
\begin{equation}\label{flow}
F_{\rm max}\approx (\langle\Delta p_x\rangle_{c})
/y_{\rm cm}+(\int_{0}^{t_r}F_x dt)/y_{\rm cm}.
\end{equation}
In the above, $\langle \Delta p_x\rangle_c$ is the net transverse momentum 
generated by nucleon-nucleon collisions; $F_x$ is the nuclear force acting on 
nucleons along the $+x$ direction in the reaction plane; and
$t_r\propto 1/\gamma$ is the reaction time. In pure cascade calculations,
the flow is due to nucleon-nucleon collisions, which decreases 
with increasing beam energy as a result of a rapidly decreasing 
reaction time $t_r$ during which the thermal pressure creates a sideward 
deflection in the reaction plane. Including mean-field potential in the model
leads to the second term, which depends not only on the reaction time but
also on the density gradient of the compressed matter. Although the latter 
increases with increasing beam energy, which gives rise to a stronger 
nuclear force at high energies, the shorter reaction time at higher 
energies reduces its effect. One thus expects that the second term in Eq.
(\ref{flow}) contributes more or less equally at all energies, and this
is indeed seen from the near constant difference between the two 
results shown in the figure.

\vspace*{9cm}
\includegraphics{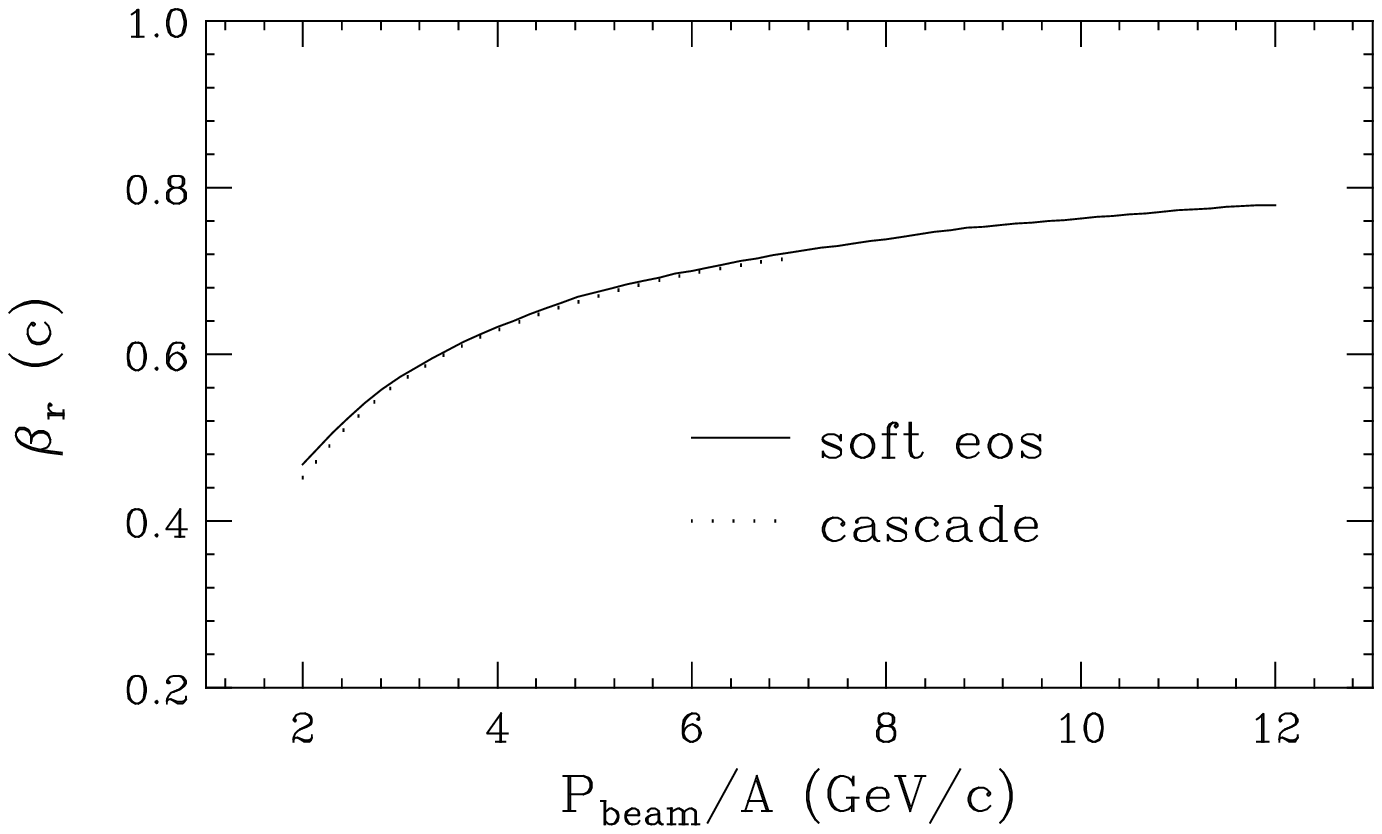}\label{bali-fig5}
\begin{minipage}[t]{6in}
\noindent Figure 5. 
The average radial flow velocity as a function
of beam momentum for the reaction of Au+Au at an impact
parameter of 2 fm. 
\end{minipage}
\vskip 1cm

The radial flow is characterized by the average radial 
flow velocity $\beta_r$ defined by 
\begin{equation}
\beta_r\equiv \frac{1}{N}\sum_i^{N}\frac{\vec{p_i}}
{E_i}\cdot \frac{\vec{r_i}}{r_i},
\end{equation} 
where the summation is over all test particles in the system.
The value of $\beta_r$ at t=20 fm/c is shown in Fig. 5 
for both the cascade and the transport model with the soft equation 
of state. We see that the radial flow velocity increases quickly with
the beam momentum and reaches a constant value of about 0.78 c. The similar 
results from these two cases further indicate that the radial flow is mainly 
determined by the thermal pressure rather than the potential pressure. 
Since the thermal pressure increases with increasing beam energy, 
the radial flow also becomes larger at higher energies.

\section{SUMMARY}

Based on the relativistic transport model (ART 1.0) 
we have studied the excitation functions of stopping, compression 
and collective flow in central Au+Au reactions at beam momenta 
from 2 to 12 GeV/c per nucleon. We find that complete stopping is reached
in the whole beam momentum range and the proton rapidity distribution scales 
with the beam rapidity. A large volume of superdense hadronic matter 
is seen to form in these reactions.
The calculated maximum baryon and energy densities are in good agreement 
with predictions of the relativistic hydrodynamics which assumes the formation 
of a shock wave in heavy-ion collisions. The nucleon transverse flow turns out
to be sensitive to the equation of state in this beam momentum range, 
and its strength is found to decrease with increasing beam momentum. 
On the other hand, the radial flow is insensitive to the nuclear equation
of state and increases with the beam momentum. In our model, neither the
quark-gluon plasma nor the chiral restored phase is included. The 
results presented here are based on purely hadronic dynamics. They are,
however, useful in the sense that deviations of experimental observations
from our predictions might signal the existence of new phenomena that 
are related to chiral symmetry restoration 
and/or quark-gluon-plasma formation in heavy-ion collisions at 
Brookhaven's AGS.

This work was supported in part by the NSF Grant No. PHY-9509266.


\begin{thebibliography}{9}

\bibitem{gyu}M. Gyulassy, Nucl. Phys. A590, 431c (1995).

\bibitem{rischke}D.H. Rischke, in Proc. of Heavy-Ion Physics at the 
AGS, edited by C.A. Pruneau {\it et al.}, p. 138 (1996).

\bibitem{art1}B.A. Li and C.M. Ko, Phys. Rev. C52, 2037 (1995); 
{\it ibid}, C53, R22 (1996).

\bibitem{e866n}F. Videbaek, Nucl. Phys. A590, 249c (1995).

\bibitem{liko96}B.A. Li and C.M. Ko, Nucl. Phys. A601, 457 (1996). 

\bibitem{li91a}B.A. Li and W. Bauer, Phys. Lett. B254, 335 (1991); 
        Phys. Rev. C44, 450 (1991).
 
\bibitem{harris}J.W. Harris, in Advances in Nuclear Dynamics 2, 
	Eds. W. Bauer and G.D. Westfall, Plenum Press, New York, 1996, 
        page 401.

\bibitem{fl}E. Fermi, Prog. Theor. Phys. V5, 750 (1950);
L.D. Landau, Izv. Akad. Nauk SSSR, Ser. Fiz. V17, 51 (1953).  

\bibitem{wong95}C.Y. Wong, Introduction to High Energy Heavy-Ion Collisions
(World Scientific, Singapore, 1994).

\bibitem{taub}A.H. Taub, Phys. Rev. 74, 328 (1948).

\bibitem{dan}R.B. Clare and D. Strottman, 
Phys. Rep. 141, 177 (1986).

\bibitem{nix}A.A. Amsden, F.H. Harlow and J.R. Nix, 
	Phys. Rev. C15, 2059 (1977).

\bibitem{dani}P. Danielewicz and G. Odyniec, Phys. Lett. B157, 146 (1985).

\end{thebibliography}
\end{document}